\documentclass[twocolumn,showpacs,prl]{revtex4}
\usepackage{graphicx}
\usepackage{dcolumn}
\usepackage{bm}

\begin{document}

\title{Spin Hall effects for cold atoms in a light induced gauge potential}
\author{Shi-Liang Zhu$^{1,2}$, Hao Fu$^{1}$, C.-J. Wu$^{3}$, S. -C. Zhang$^{4}$, and
L. -M. Duan$^{1}$}
\address{$^1$FOCUS center and MCTP, Department of Physics, University of
Michigan, Ann Arbor, MI 48109\\
$2$Institute for Condensed Matter Physics and SPTE, South China
Normal University, Guangzhou, China\\ $^3$ Kavli Institute for
Theoretical Physics, University of
California, Santa Barbara, CA 93106\\
$^4$ Department of Physics, Stanford University, Stanford, CA
94305-4045}

\begin{abstract}
We propose an experimental scheme to observe spin Hall effects with cold
atoms in a light induced gauge potential. Under an appropriate
configuration, the cold atoms moving in a spatially varying laser field
experience an effective spin-dependent gauge potential. Through numerical
simulation, we demonstrate that such a gauge field leads to observable spin
Hall currents under realistic conditions. We also discuss the quantum spin
Hall state 
in an optical lattice.
\end{abstract}

\maketitle

The spin Hall effect has recently attracted strong interest in condensed
matter physics because of 
its connection to quantum Hall physics \cite{Klitzing, Tsui} and 
its potential applications in spintronics \cite
{Hirsch,Murakami,Sinova,Kato,Kane,Hu,Bernevig}. In analogy to the
conventional Hall effect related to charge currents, the spin Hall effect
refers to the generation of a spin current transverse to an applied electric
field. It has been proposed to occur in certain solid-state systems with
some primitive experimental demonstration \cite{Kato}. An essential
requirement for observation of the spin Hall effect is generation of an
effective spin-dependent gauge potential either in momentum space \cite
{Murakami,Sinova,Kane} or in real space \cite{Bernevig}. In both cases,
spin-orbit coupling in semiconductors or graphene are employed to provide
such mechanisms. 

It has been widely acknowledged that ultracold atomic gases provide an ideal
playground to experimentally investigate some fundamental phenomena
originally connected with condensed matter systems \cite{ketterle}. The
remarkable controllability in 
these systems allows a clean study of many complicated physics in a
controllable fashion. The generation of an effective gauge potential in
atomic systems has raised significant interest, from the earlier
implementation of rotating traps \cite{Ho} to the more recent work on light
induced gauge fields \cite
{Jaksch,Sorensen,Juzeliunas,Juzeliunas2006,Osterloh,Dum,Raithel}. Although
most previous work focuses on the study of scalar gauge fields, it is
natural to ask whether it is possible to study \textit{spin} Hall effect in
an atomic system.

In this paper, we propose an experimental scheme for observation of the spin
Hall effects in a cold atomic gas. We show that for atoms with a simple $%
\Lambda $-type three-level configuration moving in a spatially varying laser
field, a spin-dependent gauge potential in the real space naturally arises
in connection with the Berry phase associated with the atomic motion. Under
an applied effective ``electric'' field, which can be generated from gravity
for instance, the atoms will follow a spin-dependent trajectory, which leads
to a net spin current in the direction perpendicular to the ``electric'' and
the gauge field while the mass current is zero. Furthermore, we show it is
easy to generate different forms of the gauge field in this system, with a
strong periodic gauge field as an example. The diverse configurations of the
gauge field, in combination with the tunable interaction and the
controllable potentials for the atomic gas, may allow us to study various
kinds of interesting Hall physics in this system. With this gauge field, we
also discuss the associated quantum spin Hall effect for fermionic atoms in
an optical lattice.

We consider an atomic gas with each atom having an $\Lambda $-type level
configuration as shown in Fig. 1a. The ground states $|1\rangle $ and $%
|2\rangle $ are coupled to an excited state $|3\rangle $ through spatially
varying laser fields, with the corresponding Rabi frequencies $\Omega _{1}$
and $\Omega _{2}$, respectively. Different from the previous work \cite
{Juzeliunas,Juzeliunas2006}, we assume here off-resonant couplings for the
single-photon transitions with the same large detuning $\Delta $, and we
will use the bright state as well as the dark state to realize a spin
dependent gauge field for the atoms.

The full quantum state of the atoms $|\Phi (\mathbf{r})\rangle $ (including
both the internal and the motional degrees of freedom) can be expanded as $%
|\Phi (\mathbf{r})\rangle =\sum_{j=1}^{3}\phi _{j}(\mathbf{r})|j\rangle $,
where $\mathbf{r}$\ denotes the atomic position. The Hamiltonian of the atom
has the form $H=\frac{\mathbf{P}^{2}}{2m}+V(\mathbf{r})+H_{int}$, where $m$
is the atomic mass, $V(\mathbf{r})$ denotes the external trapping potential
which we assume to be diagonal in the internal states $|j\rangle $ with the
form $V(\mathbf{r})=\sum_{j}V_{j}(\mathbf{r})|j\rangle \langle j|$ , and $%
H_{int}$ is the laser-atom interaction Hamiltonian, given by
\begin{equation}
H_{int}=\left(
\begin{array}{lll}
0 & 0 & \Omega _{1} \\
0 & 0 & \Omega _{2} \\
\Omega _{1}^{\ast } & \Omega _{2}^{\ast } & 2\Delta
\end{array}
\right)  \label{H_int}
\end{equation}
in the basis $\left\{ |1\rangle ,|2\rangle ,|3\rangle \right\} $. We
parameterize the Rabi frequencies through $\Omega _{1}=\Omega \sin \theta
e^{i\varphi }$ and $\Omega _{2}=\Omega \cos \theta $ with $\Omega =\sqrt{%
|\Omega _{1}|^{2}+|\Omega _{2}|^{2}}$ ($\theta $ and $\varphi $ are in
general spatially varying). The eigenvectors (the dressed states) $|\chi
\rangle =(|\chi _{1}\rangle ,|\chi _{2}\rangle ,|\chi _{3}\rangle )^{Tr}$ of
the Hamiltonian $H_{int}$ are specified by $|\chi \rangle =U(|1\rangle
,|2\rangle ,|3\rangle )^{Tr}$ ($Tr$ denotes the transposition), where
\begin{equation}
U=\left(
\begin{array}{lll}
\cos \theta & -\sin \theta e^{-i\varphi } & 0 \\
\sin \theta \cos \gamma e^{i\varphi } & \cos \theta \cos \gamma & -\sin
\gamma \\
\sin \theta \sin \gamma e^{i\varphi } & \cos \theta \sin \gamma & \cos \gamma
\end{array}
\right) ,  \label{U}
\end{equation}
and $\gamma $ is given by $\tan \gamma =(\sqrt{\Delta ^{2}+\Omega ^{2}}%
-\Delta )/\Omega ,$ with the corresponding eigenvalues $\lambda =(0,\Delta -%
\sqrt{\Delta ^{2}+\Omega ^{2}},\Delta +\sqrt{\Delta ^{2}+\Omega ^{2}})^{Tr}.$
In the new basis $|\chi \rangle $, the full quantum state of the atom $|\Phi
(\mathbf{r})\rangle $ is written as $|\Phi (\mathbf{r})\rangle =\sum_{j}\Psi
_{j}(\mathbf{r})|\chi _{j}(\mathbf{r})\rangle $, where the wave functions $%
\Psi =(\Psi _{1},\Psi _{2},\Psi _{3})^{Tr}$ obey the Schr\"{o}dinger
equation $i\hbar \partial _{t}\Psi =\tilde{H}\Psi $, with the effective
Hamiltonian $\tilde{H}$ taking the form:
\begin{equation}
\tilde{H}=\frac{1}{2m}(-i\hbar \nabla -\mathbf{\tilde{A}})^{2}+\tilde{V}(%
\mathbf{r}),  \label{H_1}
\end{equation}
where $\mathbf{\tilde{A}}=i\hbar U\nabla U^{\dagger }$ and $\tilde{V}(%
\mathbf{r})=\lambda I+UV(\mathbf{r})U^{\dagger }$ ($I$ is the $3\times 3$
unit matrix )\cite{Wilczek}. From Eq.(\ref{H_1}), one can see that in the
new basis the atoms can be considered as moving in a gauge potential $%
\mathbf{\tilde{A}}$ and a scalar potential $\tilde{V}(\mathbf{r})$.

\begin{figure}[tbph]
\label{Fig1}
\includegraphics[height=4cm,width=8cm]{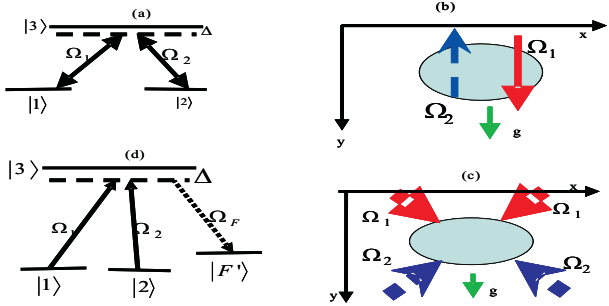}
\caption{ (Color online) Schematic representation of the
light-atom interaction for generation of effective spin-dependent
gauge fields. (a) Three-level $\Lambda$-type atoms interacting
with laser beams characterized by the Rabi frequencies $\Omega_1$,
$\Omega_2$ through the Raman-type coupling with a large
single-photon detuning $\Delta$. (b)The Configurations of the
Raman laser beams. Configuration $I$: two counter-propagating and
overlapping laser beams with shifted spatial profiles (see also
Ref. \protect\cite{Juzeliunas2006}). (c) Configuration $II$: A
periodic gauge field can be created by four overlapping laser
beams propagating along the shown directions. The upper two form
the Raman beam $\Omega_1$ while the lower two form $\Omega_2$. (d)
A Raman configuration to transfer the bright state to a different
hyperfine level $|F'\rangle$ for detection. The $|1\rangle$ and
$|2\rangle$ are assumed to be different Zeeman states on the same
hyperfine level $|F\rangle$.}
\end{figure}

We are interested in the subspace spanned by the two lower internal
eigenstates $\left\{ |\chi _{1}\rangle ,|\chi _{2}\rangle \right\} $ (called
respectively the dark and the bright state). This gives an effective
spin-1/2 system, and in the spin language we also denote $|\chi _{\uparrow
}\rangle \equiv |\chi _{1}\rangle $ and $|\chi _{\downarrow }\rangle \equiv
|\chi _{2}\rangle $. In the case of a large detuning ($\Delta >>\Omega $),
both states $|\chi _{\uparrow }\rangle $ and $|\chi _{\downarrow }\rangle $
have negligible contribution from the initial excited-state $|3\rangle $, so
they are stable under atomic spontaneous emission. Furthermore, we assume
the adiabatic condition, which requires the off-diagonal elements of the
matrices $\mathbf{\tilde{A}}$ and $\tilde{V}$ are much smaller than the
eigenenergy differences $\left| \lambda _{i}-\lambda _{j}\right| $ ($%
i,j=1,2,3$) of the states $|\chi _{i}\rangle $. This gives the quantitative
condition $F\ll \Omega ^{2}/2\Delta $, where $F=\cos ^{2}\theta |\mathbf{v}%
\cdot \nabla (\tan \theta e^{i\varphi })|$ ($\mathbf{v}$ is the typical
velocity of the atom) represents the two-photon Doppler detuning\cite
{Juzeliunas2006}. Under this adiabatic condition, the Schrodinger equation
for the wave function $\Psi $ becomes diagonal in the basis $\left\{ |\chi
_{i}\rangle \right\} $, and in the lower subspace spanned by $\left\{ |\chi
_{\uparrow }\rangle ,|\chi _{\downarrow }\rangle \right\} $, the effective
Hamiltonian takes the form
\begin{equation}
H_{eff}=\left(
\begin{array}{cc}
H_{\uparrow } & 0 \\
0 & H_{\downarrow }
\end{array}
\right) ,  \label{H}
\end{equation}
where $H_{\sigma }=\frac{1}{2m}(-i\hbar \nabla -\mathbf{A}_{\sigma
})^{2}+V_{\sigma }(\mathbf{r}),$ $\left( \sigma =\uparrow ,\downarrow
\right) $. The gauge and the scalar potentials $\mathbf{A}_{\sigma }$ and $%
V_{\sigma }$ for the spin-$\sigma $ component are given by $\mathbf{A}%
_{\sigma }=i\hbar \langle \chi _{\sigma }|\nabla |\chi _{\sigma
}\rangle $ and $V_{\sigma }(\mathbf{r})=\lambda _{\sigma }+\langle
\chi _{\sigma }|V|\chi _{\sigma }\rangle + \frac{\hbar
^{2}}{2m}[\langle \nabla \chi _{\sigma }|\nabla \chi _{\sigma
}\rangle +|\langle \chi _{\sigma }|\nabla \chi _{\sigma }\rangle
|^{2}] $, respectively. Through Eq. (2), one can find out that
$\mathbf{A}_{\uparrow }=-\mathbf{A}_{\downarrow }=-\hbar \sin
^{2}\theta \nabla \varphi $ and the related gauge field
\begin{equation}
\mathbf{B}_{\sigma }=\nabla \times \mathbf{A}_{\sigma }=-\eta _{\sigma
}\hbar \sin (2\theta )\nabla \theta \times \nabla \varphi ,  \label{B}
\end{equation}
where $\eta _{\uparrow }=-\eta _{\downarrow }=1$. We get exactly a
spin-dependent gauge field from the above configuration of the laser-atom
coupling, which is critical for the \textit{spin} Hall effect.

We consider two specific configurations of the laser beams, which generate
different spatial variations of the gauge field. First, two
counter-propagating Gaussian laser beams with shifted centers generate a
spatially slowly varying gauge field \cite{Juzeliunas2006}. The spatial
profiles of the corresponding Rabi frequencies $\Omega _{j}$ have the form $%
\Omega _{j}=\Omega _{0}\exp [-(x-x_{j})^{2}/\sigma _{0}^{2}]\exp (-ik_{j}y),$
$\left( j=1,2\right) $, where the propagating wave vectors $k_{1}=-k_{2}=k/2$
and the center positions $x_{1}=-x_{2}=\Delta x/2$ (see Fig. 1b). Under
these two laser beams, the gauge field is given by
\begin{equation}
\mathbf{A}_{\sigma }^{I}=\frac{-\eta _{\sigma }\hbar k}{1+e^{-x/d}}\mathbf{e}%
_{y},\text{ }\ \mathbf{B}_{\sigma }^{I}=\frac{\eta _{\sigma }\hbar k}{%
4d\cosh ^{2}(x/2d)}\mathbf{e}_{z},  \label{Field1}
\end{equation}
where $d=\sigma _{0}^{2}/\left( 4\Delta x\right) $. Second, through
overlapping of two standing-wave laser beams as shown in Fig. 1c with the
corresponding Rabi frequencies $\Omega _{1}=\Omega _{0}\cos \left( kx\sin
\alpha \right) e^{iky\cos \alpha }$ and $\Omega _{2}=\Omega _{0}\sin \left(
kx\sin \alpha \right) e^{-iky\cos \alpha }$ ($\alpha $ is the angle of the
propagating laser beams to the $y$ axis), we generate a spatially periodic
gauge field, given by
\begin{equation}
\ \mathbf{A}_{\sigma }^{II}=2\eta _{\sigma }\hbar k^{\prime }\sin
^{2}(k^{\prime }x)\mathbf{e}_{y},\ \mathbf{B}_{\sigma }^{II}=2\eta _{\sigma
}\hbar {k^{\prime }}^{2}\sin (2k^{\prime }x)\mathbf{e}_{z}  \label{Field2}
\end{equation}
with $k^{\prime }=k\sin \alpha $. Under a slowly varying gauge field $%
\mathbf{B}_{\sigma }^{I}$, one can have local Landau levels, and with a
spatially periodic gauge field $\mathbf{B}_{\sigma }^{II}$, we expect to
have Bloch-type of wave functions, similar to the case of particles in a
periodic potential.

The spin Hall effect can be demonstrated by observing a spin-Hall current.
In the following, first we propose an experiment to detect the spin Hall
current in an atomic gas with the above configuration of the light-atom
coupling, and then we discuss the quantum spin Hall effect in an optical
lattice. For observation of the spin Hall effect, we need an effective
``electric'' field $\mathbf{E}$ which drives atoms in one direction, and a
spin current should be observed in a direction perpendicular both to the
``electric'' field $\mathbf{E}$ and the gauge field $\mathbf{B}_{\sigma }$.
The ``electric'' field can be conveniently provided through gravity on the
neutral atoms. We assume the internal state of the atoms is in superposition
of the spin $\uparrow $ and $\downarrow $ components under the laser beams
shown in Fig. 1b or 1c. The external atomic trap is turned off at time $t=0$%
, and the atoms fall off due to gravity with an acceleration
$g=9.8m/s^{2}$ (along the direction $\mathbf{e}_{y}$). Under the
effective gauge field, the equations of motion are
\begin{equation}
\dot{x}_{\sigma }=p_{\sigma }^{x}/m,\ \ \dot{p}_{\sigma }^{x}=[
(\partial _{x}A_{\sigma })p_{\sigma }^{y}-A_{\sigma }\partial _{x}A_{\sigma }%
]/m-\partial _{x}V_{\sigma },  \label{Trajectory1}
\end{equation}
\begin{equation}
\dot{y}_{\sigma }=[p_{\sigma }^{y}-A_{\sigma }]/m,\ \
\dot{p}_{\sigma }^{y}=mg,  \label{Trajectory2}
\end{equation}
where the gauge potential $A_{\sigma }$ is either $A_{\sigma }^{I}$ or $%
A_{\sigma }^{II}$, and $V_{\sigma }$ is the corresponding scalar potential
induced by the same laser beams. The coordinates and the momenta $x_{\sigma
},y_{\sigma },p_{\sigma }^{x},p_{\sigma }^{y}$ are understood as variables
(operators) in the classical (quantum) cases, respectively.

\begin{figure}[tbph]
\vspace{-1.0cm} \label{Fig2}
\includegraphics[height=10.0cm]{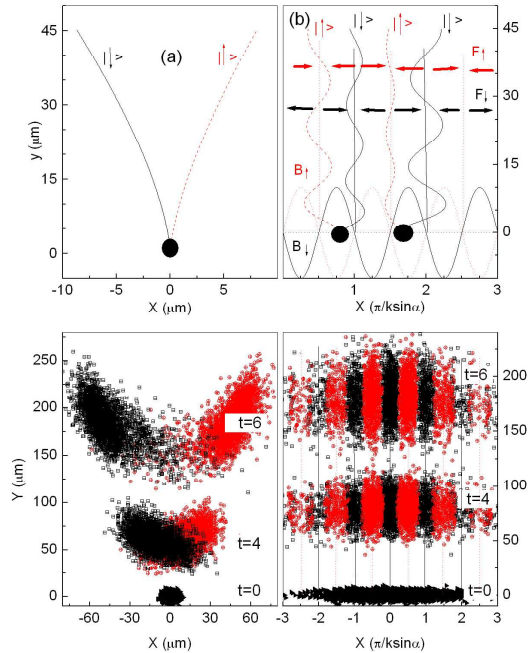}
\caption{(Color online) Spin-dependent trajectories of a single atom (a and
b) and spin-dependent evolution of the density profiles of an ensemble of
atomic gas (c and d) under gravity (which provides an effective electric
filed) and a light-induced gauge potential. The spin current along the $x$
direction is a manifestation of the spin Hall effect. The gauge potentials
in Figs. (a,c) and (b,d) are generated by the laser configurations I and II,
respectively. The sinusoids in Fig. (b) denote the effective gauge fields $%
B_{\protect\sigma}$. The directions of the Lorentz forces $F_{\protect\sigma}
$ change periodically in this case and are shown by the arrows there. The
dotted (solid) vertical lines in (b) and (d) denote the stable equilibrium
positions for spin-up (spin-down) atoms. In Figs. (c) and (d), the density
profiles of the atomic gas are shown at time $t=0,4,6\ ms$. For calculations
in Figs. (a-d), we take the following typical experimental parameters with $%
\protect\sigma_0=10\ \protect\mu m$, $\Delta x=2.5\ \protect\mu m$, $k=10^7\
m^{-1}$ for the laser configuration I, and $k\sin\protect\alpha=5 \times
10^{5}\ m^{-1}$ for the laser configuration II. In both configurations, $%
\Omega_0^2/\Delta=10^6$ Hz. In Figs. (a) and (b), the initial atomic
velocity is assumed to be zero, and the initial positions $x=0,y=0$ for (a)
and ($x=2.5,5\protect\mu m$, y=0) for (b). The parameters for the atomic
ensemble in Figs. (c) and (d) are given by $\protect\sigma_r=2.0\ \protect\mu
m$ and $\protect\sigma_v=0.5\ cm/s$. The atomic mass is taken to be the one
for $^{87}$Rb. With the above parameters, we have checked the adiabatic
condition is well satisfied during the evolution.}
\end{figure}

To have some intuitive idea, in Fig. 2 (a) and (b) we show the typical
classical trajectories of the atoms under the gauge fields $B_{\sigma }^{I}$
or $B_{\sigma }^{II}$. One can clearly see that the trajectory of the atom
depends on its spin state $\sigma $, and such a dependence leads to the spin
Hall current in the horizontal direction for the case of many particles. For
the gauge field $B_{\sigma }^{I}$, the trajectory is a parabola, while for $%
B_{\sigma }^{II}$ it is an oscillation around the nearest stable point. The
spin-dependent stable points for $B_{\sigma }^{II}$ are determined by the
zeros of the corresponding Lorentz force, which are given by $%
x_{n}=(n+1/2)\pi /k\sin\alpha$ ($x_n=n\pi /k\sin\alpha$) with an integer $n$
for the spin-$\uparrow $ ($\downarrow $) component, respectively.

The trajectory of a single atom is hard to detect, and it is much easier in
experiments to measure the density evolution of an ensemble of
non-interacting atoms. We assume at $t=0$ (the moment when the trap is
turned off), the number density and the velocity distribution of the atomic
gas are both described by Gaussian functions with $\rho _{r}(x,y)=(2\pi
\sigma _{r}^{2})^{-1}e^{-(x^{2}+y^{2})/2\sigma _{r}^{2}}$ and $\rho
_{v}(v_{x},v_{y})=(2\pi \sigma _{v})^{-1}e^{-(v_{x}^{2}+v_{y}^{2})/2\sigma
_{v}^{2}}$, respectively, where $\sigma _{r}$ and $\sigma _{v}$ characterize
the corresponding spatial and velocity variances. These variances include
contributions from both quantum uncertainties of the atomic motion and
classical broadening due to the finite temperature effect. The evolution of
the density profile of the atomic gas is simulated numerically by solving
Eqs. (\ref{Trajectory1}) and (\ref{Trajectory2}), and the results are shown
in Fig. 2 (c) and 2(d) for the gauge fields $B_{\sigma }^{I}$ and $B_{\sigma
}^{II}$ respectively. Under $B_{\sigma }^{I}$, the ensemble splits into the
spin-up and spin-down clusters, which is a manifestation of the spin Hall
current along the $x$ direction. Under $B_{\sigma }^{II}$, the atoms form
periodic patterns with micro-separation of the different spin components.

To experimentally detect the spin current (or spin separation) as shown in
Fig.2, right before the imaging one can transfer the dressed bright state $%
|\chi _{\downarrow }\rangle $ to a different hyperfine level $|F^{\prime
}\rangle $ by turning on a laser pulse (with a Rabi frequency $\Omega _{F}$)
that couples the excited state $|3\rangle $  to $|F^{\prime }\rangle $ (see
Fig. 1d). This pulse, together with the original laser beams $\Omega _{1}$
and $\Omega _{2}$, make a Raman transition with an effective Hamiltonian $%
H_{R}=\left( \Omega _{F}^{\ast }\Omega /\Delta \right) |\chi _{\downarrow
}\rangle \left\langle F^{\prime }\right| +h.c.$ (note that the dark state $%
|\chi _{\uparrow }\rangle $ is still decoupled because of the phase relation
between $\Omega _{1}$ and $\Omega _{2}$). Although the form of the bright
state $|\chi _{\downarrow }\rangle $ is spatially varying, the Rabi
frequency $\Omega $ (and thus also $\Omega _{F}^{\ast }\Omega /\Delta $) is
spatially constant (for the laser configuration II) or almost constant (in
the overlap region for the laser configuration I). We can thus choose the
pulse duration so that it makes a complete Raman transition ($\pi $-pulse),
and the atomic motion can be neglected during such a short duration. After
this Raman $\pi $-pulse, the initial different dressed spin states are
mapped to different hyperfine levels, and the populations in different
atomic hyperfine levels can be separately imaged with the known experimental
techniques.

We now consider quantum spin Hall effect with fermionic atoms in
an optical lattice. In this case, the scalar potential $V_{\sigma
}(\mathbf{r})$ is spatially periodic. We assume the optical
lattice has a higher
intensity along the vertical direction so that the tunneling rate along the $%
z$-axis is negligible. One then has an effective 2D system in the
$x-y$ plane. The gauge field $\mathbf{B}_{\sigma }\left(
\mathbf{r}\right) $ (along the $z$-axis) is assumed to nearly
constant or spatially periodic in the lattice (which corresponds
to the above laser configurations I and II, respectively). The
wave function in this case can
still be written as $\Psi _{\sigma }\left( \mathbf{r}\right) =\sum_{n\mathbf{%
k}}u_{n}^{\sigma }(\mathbf{k,r})e^{i\mathbf{k\cdot r}}$, where $\mathbf{k}$
is the Bloch wave-vector and the $n$-th band wavefunction $u_{n\sigma }(%
\mathbf{k,r})$ satisfies the Schrodinger equation with the
effective Hamiltonian\cite{Thouless}
\begin{equation}
H_{\mathbf{k}}^{\sigma }=\frac{\hbar ^{2}}{2m}\{(-i\partial
_{x}+k_{x})^{2}+[-i\partial _{y}+k_{y}-A_{\sigma
}(x)]^{2}\}+V_{\sigma }(x,y) \label{H_k}
\end{equation}
Under an effective ``electric'' field $\mathbf{E}$ along the $y$-direction ($%
E_{y}=mg$ through acceleration $g$), the Hall current along the $x$
direction is given by $J_{x}^{\sigma }\equiv \left\langle \rho _{n}^{\sigma
}v_{x}^{\sigma }\right\rangle =\sigma _{xy}^{\sigma }E_{y}$ for the spin-$%
\sigma $ component with the linear response theory. The Hall conductivity
then has the expression $\sigma _{xy}^{\sigma }=\left( 1/2\pi \hbar \right)
\sum_{n,\mathbf{k}}\rho _{\sigma }(\epsilon _{n}^{\sigma }(\mathbf{k}%
))\left( \partial _{k_{x}}a_{ny}^{\sigma }-\partial _{k_{y}}a_{nx}^{\sigma
}\right) ,$ where $a_{n\mu }^{\sigma }(\mathbf{k})\equiv i\hbar \langle
u_{n}^{\sigma }(\mathbf{k})|\partial _{k_{\mu }}u_{n}^{\sigma }(\mathbf{k}%
)\rangle $ ($\mu =x,y$) and $\rho ^{\sigma }(\epsilon _{n}^{\sigma }(\mathbf{%
k}))$ denote the density of states of the n-th band with \ the band energy $%
\epsilon _{n}^{\sigma }(\mathbf{k})$. The mass and the spin currents are
defined by $J_{x}^{m}=J_{x}^{\uparrow }+J_{x}^{\downarrow }$ and $%
J_{x}^{s}=J_{x}^{\uparrow }-J_{x}^{\downarrow }$, respectively. With $%
A_{\uparrow }=-A_{\downarrow }$ and $V_{\sigma }$ nearly independent of the
spin $\sigma $, we have $J_{x}^{m}=0$ and $J_{x}^{s}=2J_{x}^{\uparrow }$.
So, there is a net spin current, as a characteristic feature of the spin
Hall effect. The spin Hall current is quantized if the chemical potential of
the system (controlled by the atom number density) is inside a bandgap. In
this case, $\sigma _{xy}^{\sigma }=\left( 1/2\pi \hbar \right)
C_{xy}^{\sigma }$, where $C_{xy}^{\sigma }=i/2\pi \int dk_{x}dk_{y}(\langle
\partial _{k_{x}}u_{n}^{\sigma }|\partial _{k_{y}}u_{n}^{\sigma }\rangle
-\langle \partial _{k_{y}}u_{n}^{\sigma }|\partial _{k_{x}}u_{n}^{\sigma
}\rangle )$ is the Chern number which takes only integer values \cite
{Thouless}. The spin current is then an integer multiple of $1/\left( \pi
\hbar \right) $.

Finally, we briefly discuss detection of the quantum spin Hall effect. With
a nearly constant gauge field (such as the one provided by the laser
configuration I in Fig. 1), the cyclotron length scale is estimated by $%
l\sim \sqrt{\hbar /B_{I}}\sim \sigma _{0}/\sqrt{k\Delta x}$. The region of
the gauge field is characterized by the laser profile with an area about $%
S\approx 4d\times k\sigma _{0}^{2}$. The degeneracy of each Landau
level is then estimated by $S/(\pi l^{2})\sim 3.2\times 10^{3}.$
With about $10^{6}$ atoms in a three-dimensional optical lattice
which separates the atomic gas into about ${10}^{2}$ independent
layers, the atom number of each layer is of the order of $10^{4}$.
So only a few lowest Landau levels will be occupied. As the
filling number is of the order of unity, the quantum effect should
become important at low temperature. To detect the quantum spin
Hall effect, one can apply an effective ``electric'' field by
tilting the lattice along one direction (through gravity or
through lattice acceleration induced with a non-linear frequency
chirp on the laser fields that form the optical lattice), and then
detect the spin current accumulation along the vertical direction
through separate imaging of the two different spin components as
described before.

In summary, we have proposed an experimental scheme to realize effective
spin-dependent gauge potentials on cold atoms with laser beams and to
observe the spin Hall effect with an ensemble of atomic gas. Quantization of
the spin Hall effect in an optical lattice is also discussed. For ultracold
atoms with a larger density, one should also take into account of the atomic
interaction. Rich physics can arise from a combination of the proposed
spin-dependent gauge potential and the controllable atomic interaction,
which is an interesting subject for future investigation.

This work was supported by the NSF under grant numbers 0431476 and
0244841, the ARDA under ARO contracts, the A. P. Sloan Fellowship,
NSF FOCUS center and the MCTP. S.C.Z. acknowledges support by the
NSF (DMR-0342832) and the DOE (DE-AC03-76SF00515). S.L.Z. also
acknowledges support by the NCET and NSFC (10674049).

Note added: During publication of this work, we became aware that
the atomic spin Hall effect was also discussed by Liu et al. in
Ref. \cite{Liu} under a different atomic configuration

\end{document}